\documentclass[conference]{IEEEtran}
\pdfoutput=1
\IEEEoverridecommandlockouts

\usepackage[hyphens]{url}
\usepackage{hyperref}
\hypersetup{colorlinks,allcolors=blue}
\usepackage[backend=biber, style=ieee]{biblatex}

\usepackage{amsmath,amssymb,amsfonts}
\usepackage{algorithmic}
\usepackage{graphicx}
\usepackage{textcomp}
\usepackage{xcolor}
\usepackage{listings}
\usepackage{orcidlink}
\def\BibTeX{{\rm B\kern-.05em{\sc i\kern-.025em b}\kern-.08em
    T\kern-.1667em\lower.7ex\hbox{E}\kern-.125emX}}

\lstdefinestyle{bashstyle}{
    language=bash,
    breaklines=true,
    basicstyle=\ttfamily\tiny, 
    backgroundcolor=\color{lightgray!20}, 
    frame=single, 
    rulecolor=\color{gray}, 
    keywordstyle=\color{blue}, 
    morekeywords={curl, istioctl, kubectl, kg}, 
    showstringspaces=false 
}
    
\begin{document}

\title{Microsegmented Cloud Network Architecture Using Open-Source Tools for a Zero Trust Foundation\\
}

\author{\IEEEauthorblockN{Sunil Arora\,\orcidlink{0009-0007-3066-3461}}
\IEEEauthorblockA{\textit{Dakota State University}, USA \\
Sunil.Arora@trojans.dsu.edu\\ 
}
\and
\IEEEauthorblockN{John Hastings\,\orcidlink{0000-0003-0871-3622}}
\IEEEauthorblockA{\textit{Dakota State University}, USA \\
John.Hastings@dsu.edu \\
}
}

\maketitle

\begin{abstract}
This paper presents a multi-cloud networking architecture built on zero trust principles and micro-segmentation to provide secure connectivity with authentication, authorization, and encryption in transit. The proposed design includes the multi-cloud network to support a wide range of applications and workload use cases, compute resources including containers, virtual machines, and cloud-native services, including IaaS (Infrastructure as a Service (IaaS), PaaS (Platform as a service). Furthermore, open-source tools provide flexibility, agility, and independence from locking to one vendor technology. The paper provides a secure architecture with micro-segmentation and follows zero trust principles to solve multi-fold security and operational challenges. 
\end{abstract}

\begin{IEEEkeywords}
Zero trust, Networks, Cloud networking, Micro-segmentation, Service mesh, Public cloud, Multi-cloud, Application protection, Istio, Calico, Segregation, Microservices
\end{IEEEkeywords}

\section{Introduction}

Modern organizations rely on technology, applications, and computing infrastructure to deliver products and services to their customers. Traditionally, enterprises hosted computing infrastructure within on-premises data centers. However, with the adoption of contemporary technologies such as cloud computing, virtualization, Internet of Things (IoT), and Industrial Internet of Things (IIoT), applications are deployed in distributed architectures in the public cloud, hybrid cloud, or multi-cloud model. Moreover, organizations’ networks and computing infrastructure might be spread over multiple continents and countries. While this technological landscape enables companies to meet business demands and serve customer needs more effectively, it also presents heightened risks. Bad actors and hackers increasingly exploit these dispersed networks and applications to gain access to confidential data, disrupt business services, and harm organizations’ reputations. 

\begin{figure}
    \centering
    \includegraphics[width=1\linewidth]{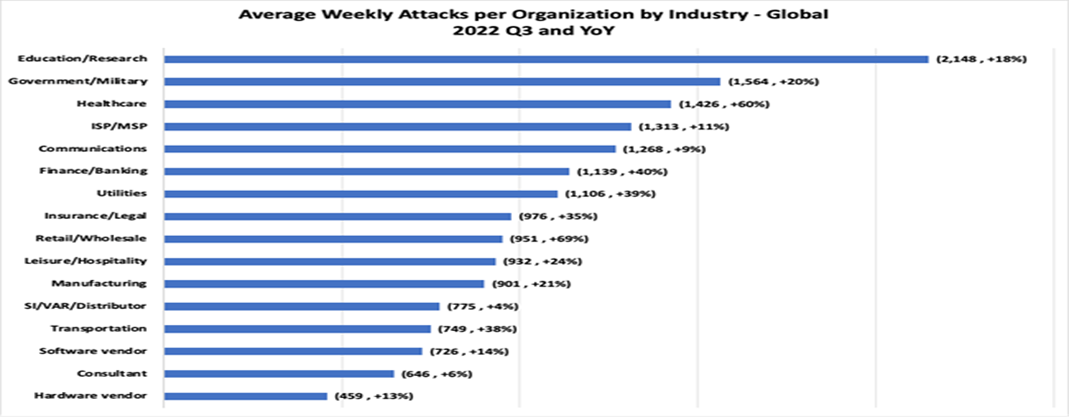}
    \caption{Average weekly attacks per organization by industry for Q3 2022. \textit{Source: \cite{1checkpoint2022}} }
    \label{fig:1}
\end{figure}

Cyberattacks have surged globally, with a 28\% year-over-year increase in incidents in the third quarter of 2022, with organizations facing over 1130 average weekly cyberattacks on their infrastructure and applications~\cite{1checkpoint2022}. As seen in Fig. \ref{fig:1}, the education and research sectors were most affected, with an average of 2,148 attacks per organization weekly during the third quarter of 2022 -- an 18\% increase compared to the same time in 2021. This reflects the significant need for appropriate protection for computing resources and applications, especially Internet-facing or hybrid networks and applications. 

Cloud computing provides a platform for modern application architecture using containers, micro-services architecture, and service mesh. These newer technologies help build flexible and agile applications and services instead of monolithic applications. The micro-service architecture divides applications into small, loosely coupled services and features with specific tasks and responsibilities. These features or services are independent of other services and can be managed, scaled, and deployed independently \cite{3gannon2017}. 

With cloud computing technology's maturity, organizations are increasingly adopting it, with the global cloud computing market 
expected to grow to USD 1240.9 billion by 2027 from USD 548.8 billion in 2022 \cite{2market2022}. 
However, traditional network security, which often emphasizes perimeter-based defenses, is no longer sufficient to address the security demands of such dynamic cloud architectures. A modern network and security design must support various applications, workloads, and environment types. It should not only filter traffic at the network layer but provide authentication, authorization, secure network paths, visibility, monitoring, and support cloud, microservices, and legacy application deployments.

This paper presents a novel multi-cloud network architecture using zero trust principles and micro-segmentation to provide secure, scalable connectivity for a variety of applications, including containers, virtual machines, and cloud-native services such as IaaS and PaaS. By utilizing open-source tools like Istio and Calico, the proposed architecture offers flexibility and agility while avoiding vendor lock-in, ensuring secure connections, authentication, and encrypted data flows across distributed environments. This architecture supports dynamic, scalable, and secure network designs across multi-cloud environments, addressing both security and operational challenges in a cost-effective manner.

The remainder of this paper is organized as follows: Section II outlines the important definitions. Section III provides an analysis of existing work and a literature review. Section IV presents the multi-cloud secure network architecture. Finally, section V outlines the use of open-source tools for service mesh to build a secure micro-service application design network. Section VI presents the architecture's benefits. Section VII includes the prototype architecture and demonstration, and Section VIII presents the paper’s conclusions and opportunities for future work.

\section{Definitions}

As per NIST Special Publication 800-207~\cite{4rose2020}, 
\textit{zero trust} and \textit{zero trust architecture} are defined as:

\begin{quote}
“Zero trust (ZT) provides a collection of concepts and ideas designed to minimize uncertainty in enforcing accurate, least privilege per-request access decisions in information systems and services in the face of a network viewed as compromised. Zero trust architecture (ZTA) is an enterprise's cybersecurity plan that utilizes zero trust concepts and encompasses component relationships, workflow planning, and access policies. Therefore, a zero trust enterprise is the network infrastructure (physical and virtual) and operational policies that are in place for an enterprise as a product of a zero trust architecture plan.”
\end{quote}

Zero trust means no trust in any device, user, computing, or connection unless validated and authorized. Authentication and authorization of each connection and session are the fundamental principles of zero trust, regardless of the client's status, location, and environment. A zero trust network assumes all network communications are threats until verified, authorized, and secured~\cite{5sheikh2021}.

Cloud networking refers to the virtualized network capabilities available by the cloud service provider for a customer's cloud computing environment. Cloud customers control the cloud virtualized or software-defined networking (SDN) design, configuration, and management as per requirements \cite{6sciarrilli2020}.

Micro-segmentation is a networking technique that divides applications and workloads into individual segments to implement granular security policies and control network traffic \cite{7vmware}.

\section{Related Work}

Several approaches have been proposed and designed for microservice architecture and segmentation; however, there are gaps and opportunities for a multi-cloud micro-segmented network architecture that supports containers, IaaS, PaaS, and cloud-native services and follows zero trust principles.

\citeauthor*{8weever2020}~\cite{8weever2020} utilized a zero-based principles approach to control and secure east-west traffic within a containerized environment. This approach uses building blocks with a container network interface (CNI) and service mesh. Similarly, \citeauthor*{9sharma2022}~\cite{9sharma2022} proposed the multi-cloud architecture using Kubernetes and Istio service mesh. This approach provides the networking components for Kubernetes and containerized applications. However, these approaches are limited to containerized applications and do not offer a method for other types of applications. These do not provide a scalable design for multi-cloud networking, micro-segmentation, and secure network architecture for various applications and workloads hosted in virtual machines, IaaS, or PaaS.

\citeauthor*{10levin2015}~\cite{10levin2015} provided an approach for cloud interoperability using the federated model; however, with the evolution of cloud services and the introduction of more cloud-native services, this approach requires building a custom federation solution and may require significant effort and development. In addition, this approach does not provide a comprehensive networking and security solution for cloud-native and micro-service architecture.

\citeauthor*{11lloret2014}~\cite{11lloret2014} proposed an Intercloud communication model that uses three layers: the organization, the distribution, and the access layers. A single Onode manages the distribution layer with Dnodes and Onodes for each cloud. It enables logical communication between different clouds but requires DNodes and Onodes in every cloud, limiting its scope of implementation in real-world scenarios. It may require significant integration efforts and be costly to deploy and operate. According to Yiliyaer and Kim \cite{12yiliyaer2022}, the following four protections should be in place in zero trust architecture. \begin{itemize}
    \item Verify every user
    \item Verify every device
    \item Enforce the least privilege
    \item Collect Information and analyze real-time
\end{itemize}
However, encryption is another vital aspect of zero trust architecture. The zero trust architecture aims to protect the data regardless of its state (at rest, in transit, or in use). Hence, encryption is a vital requirement and part of zero trust architecture \cite{13syed2022}. Micro-segmentation is another essential building block of a zero trust network. \citeauthor*{14decusatis2016}~\cite{14decusatis2016} defined the micro-segmentation of zero trust network features to include authentication of individual packets, not only the users or applications. 
Cloud SDN provides a more flexible dynamic security and governance capability. SDN can quickly adapt to changes in networking requirements and react to new threats learned in the environment \cite{15linthicum2016}. The software-defined perimeter (SDP) allows network architects and operators to deploy perimeter security functions to support the environment, servers, and applications. Logical and virtual components replace the hardware and physical appliances. The organization controls these SDP functions and features as a protection mechanism \cite{16moubayed2019}.

\section{A Modern Architecture}

\begin{figure} [h!]
    \centering
    \includegraphics[width=1\linewidth]{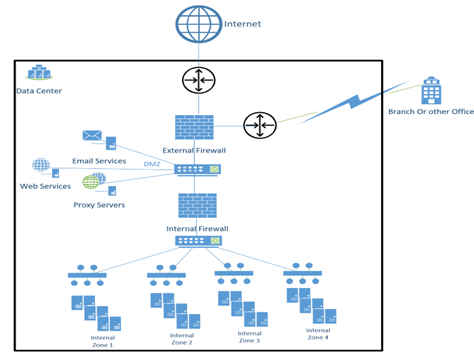}
    \caption{Traditional zoning network architecture}
    \label{fig:3}
\end{figure}

A few years back, most computer networks had flat network architecture with security controls primarily focused on the perimeter. Perimeter network security controls included firewalls, intrusion prevention systems (IPS), intrusion detection systems (IDS), virtual private networks (VPN), network proxies, web filtering, and remote access Gateways. However, most of the internal network was allowed to communicate freely. Based on the requirements, network zones, such as demilitarized zones, are used for network segmentation for Internet-facing infrastructure and applications. Perimeter network security controls were focused on OSI layer 3 and 4 access control lists to restrict the traffic by IP address and port numbers. There was no protection from advanced persistent threats and later movement once a host was compromised. A typical traditional network architecture looked like the one shown in Fig. \ref{fig:3}.

The fast development of newer technologies, including cloud, API (application programming interface), micro-services, and service-oriented architecture, has changed how applications are developed and deployed. Modern applications use agile methodology for continuous integration and deployment (CI/CD) and require rapid scaling of infrastructure components. Traditional network design and architecture can not fulfill the demand and scale to support modern application architecture. With the introduction of software defined networking (SDN), networking has become more agile and can scale and meet the demands of various application types and resources \cite{15linthicum2016}. However, a secure network architecture carefully considers evolving threats and risks. A zero trust and secure network architecture removes the inherent trust from the network while securing and validating each connection request. It requires strong authentication, authorization, encryption, and monitoring to detect and prevent cybersecurity attacks and lateral movement attempts \cite{17ncsc}.

The proposed architecture includes the following modern security concepts and technologies:
\begin{itemize}
\item Zero trust principles
\item Micro-segmentation
\item Service-mesh
\item Public cloud 
\item Native network components
\item Network perimeter
\item Network security policy management
\item Network management, including logging, monitoring and observability
\end{itemize}

\begin{figure}
    \centering
    \includegraphics[width=1\linewidth]{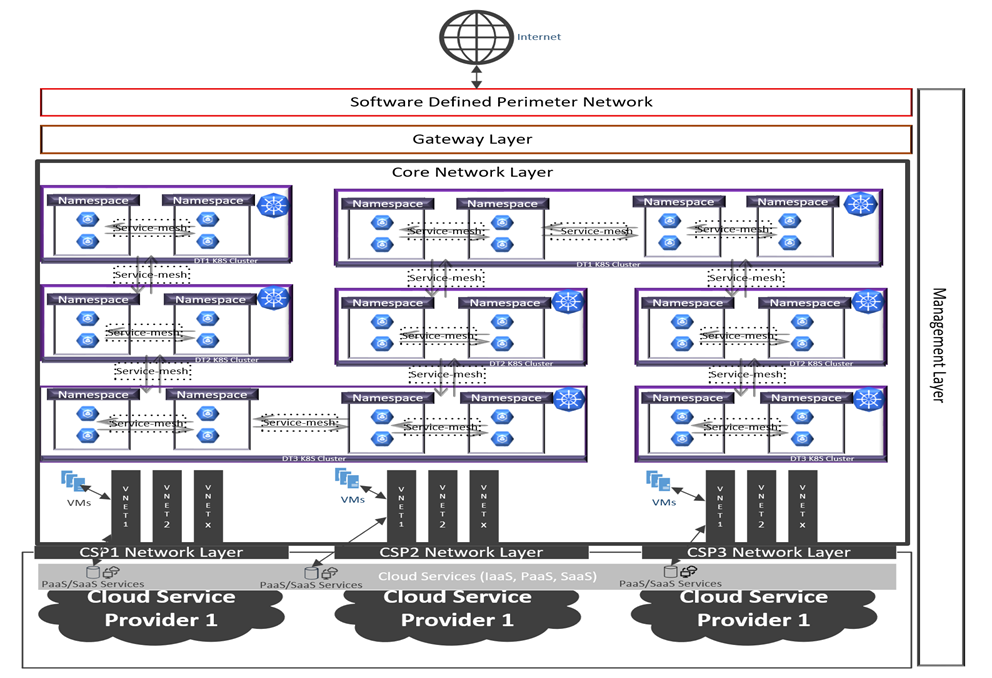}
    \caption{High-level proposed micro-segmented architecture with five layers}
    \label{fig:4}
\end{figure}

The proposed architecture includes five layers to provide a cloud-agnostic approach as highlighted in Fig. \ref{fig:4}:
\begin{enumerate}
    \item Core Network Layer
    \item Gateway Layer
    \item Software Defined Perimeter
    \item Cloud Network Layer
    \item Management Layer
\end{enumerate}
\subsection{Core Network Layer}
The Core Network Layer is the core part of the network, built based on micro-segmentation and zero trust principles. The Core Network layer contains the following components:
	
\subsubsection{Network Micro-segmentation}
	The core network layer has multiple layers of segregation.  
The first core layer contains high-level segregation based on business groups or functions. At this layer, a virtual network or group of networks segregates the business groups—for example, a virtual private cloud and subnet in Amazon Web Services \cite{18aws}. Similarly, Microsoft Azure provides subscriptions and virtual networks \cite{19azure}.

The second core layer contains the segregation of resources, such as Kubernetes or virtual machines. Virtual machines hosted in the same cloud service provider network or on-premises cloud service providers can use the same open-source tools to achieve micro-segmentation and a zero trust foundation. For example, Istio and Calico provide the VM installation agent and proxy to integrate the VM with Kubernetes \cite{20istio_vm}. The rest of the paper will refer to Kubernetes and container technologies, as these latest technologies are used to build modern architecture \cite{21kubernetes}. This segregation uses data classification, business sub-functions, or data types based on the organization's needs. For example, compute resources segmentation for low-risk, moderate-risk, and high-risk applications.

	Finally, the third core layer in that application is isolated within the security layer. For example, many open-source tools are available in a microservice architecture to build service mesh for isolation and micro-segmentation.
 
\subsubsection{Layer 3 and 4 access controls (ACLs or Security Groups)}
Network layer filtering is deployed at the OSI (Open System Interconnection) layer three network and Layer 4 transport layer. There are multiple ways to achieve this. For example, it could be achieved using cloud native capabilities such as access control lists and security groups or a software-defined overlay network such as Calico projects. Calico is an open-source network and security software that provides network policies at the host networking layer \cite{22calico}. 

\subsubsection{Application layer security controls using micro-segmentation}
This design proposes using service mesh tools to achieve application layer security that provides authentication, authorization, and encryption. Various open-source service-mesh tools exist, such as Istio or Linkered \cite{23wilson2023}. This layer also includes segregating the application resources within the Kubernetes cluster using a namespace that provides virtual segregation capability. This paper will discuss Istio features and capabilities considered in this architecture in detail in a later section.

\subsection{Gateway Layer}
The gateway layer hosts the centralized services or virtual appliances required to support the applications. For example, it may include an API gateway and a centralized encryption gateway to encrypt and decrypt certain data types as they exit and enter the core network. An API gateway intercepts client API calls and applies routing, protocol, and security policies before routing the API call to the appropriate microservice. It acts like a reverse proxy for API calls \cite{24F5Inc_APIGateway}.

\subsection{Software Defined Perimeter}
The traditional network perimeter is an old concept that protects the internal trusted network by adding a layer of security devices such as firewalls and IPS (Intrusion Prevention System). Earlier, a trusted network was hosted in on-premises data centers, and all users' machines and systems were part of the internal network. However, nowadays, users work remotely using mobile and laptops and are not restricted to the internal network. In addition, applications are deployed in various models, including IaaS, PaaS, and SaaS, and are no longer local to on-premises data centers \cite{25Sallam2019Security}. Software Defined Perimeter provides an abstraction layer for perimeter security controls using a controller and applies centralized policy to the connections, regardless of where it is deployed \cite{16moubayed2019}. Furthermore, security services may exist in the public cloud using IaaS, PaaS, or the SaaS model. Hence, route the traffic to a security service provider that can apply web filtering, TLS termination, distributed denial of services, malicious code or virus scanning, data leakage prevention, and many more security capabilities to Internet inbound and outbound traffic \cite{26Zscaler_SDP}.

\subsection{Cloud Network Layer}

The cloud network layer represents the cloud service provider's networking capabilities. Cloud service providers expose a virtual network based on software-defined networking. At this layer, the public cloud offers an abstract private virtual network that is isolated and secure from other cloud customers. It allows cloud customers to design and build a secure network as required.
This architecture uses the cloud network layer to deploy the infrastructure and compute resources in the private network. In addition, cloud service providers provide the cloud-native network components for private connectivity, such as a private link or private endpoints to connect to PaaS and SaaS without communicating via the Internet \cite{27Microsoft_AzurePrivateLink}.

\subsection{Management Layer}

The management layer contains network management and security tools and services, such as network traffic monitoring, security information and event management (SIEM), governance and audit tools, privileged access management (PAM), cryptographic key management. configuration management, and certificate management~\cite{28Kretzschmar2011Intercloud}. These tools and services may exist in the public or hybrid cloud, on-premises, or software as a service.

\section{Micro-Segmentation Using Open-Source Tools}
Various open-source tools are available to build service mesh and network security for micro-service architecture. We used the Istio service mesh to design this architecture. This section provides a detailed overview of the Istio service-mesh tool:

Network communication management and routing become very challenging as the number of services grows in a distributed architecture. Service mesh helps manage these complexities and operational requirements for service-to-service communication. Service mesh is an infrastructure layer that provides the capability to control the traffic between distributed and decoupled components transparently. Service mesh uses a proxy architecture to intercept the network traffic and apply various routing and security policies as defined \cite{29MicrosoftLearn_ServiceMeshes}. A service mesh tool delivers the following capabilities and features:

\begin{itemize}
    \item Traffic management helps to manage communication between services using the proxy. It may include routing, load balancing, inbound and outbound communication, and network failure recovery and resilience.
    \item Security policies for authentication and authorization
    \item Network traffic encryption using Mutual Transport Layer Security (mTLS)
    \item Observability includes traffic and access logging, metrics, and reporting.
\end{itemize}

\begin{figure}[h!]
    \centering
    \includegraphics[width=0.85\linewidth]{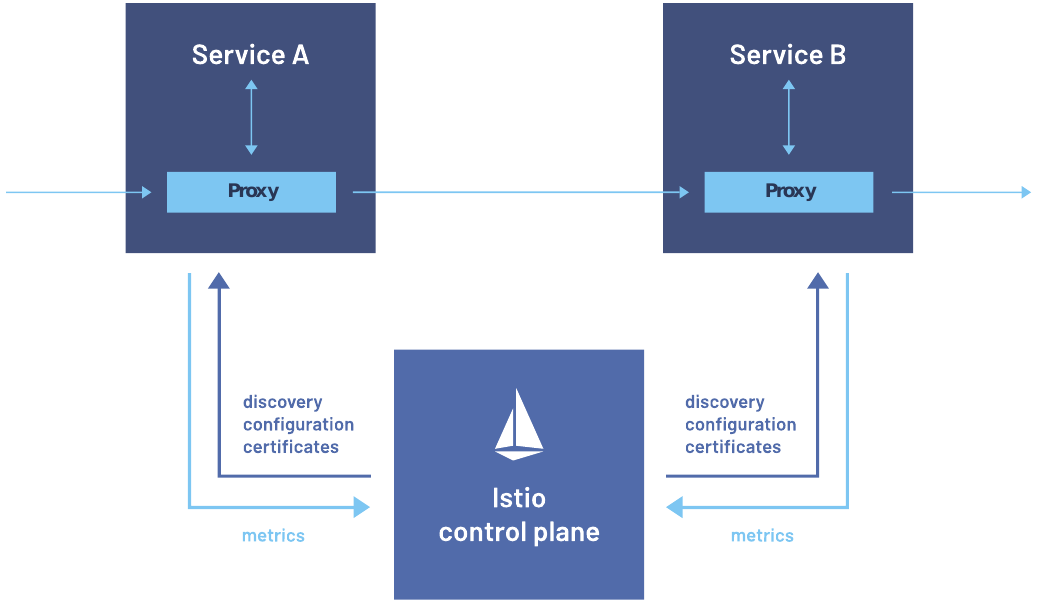}
    \caption{Istio control and data plane for communication between services A and B. \textit{Source}: \cite{30Istio_ServiceMesh}}
    \label{fig:5}
\end{figure}

\begin{figure}
    \centering
    \includegraphics[width=1\linewidth]{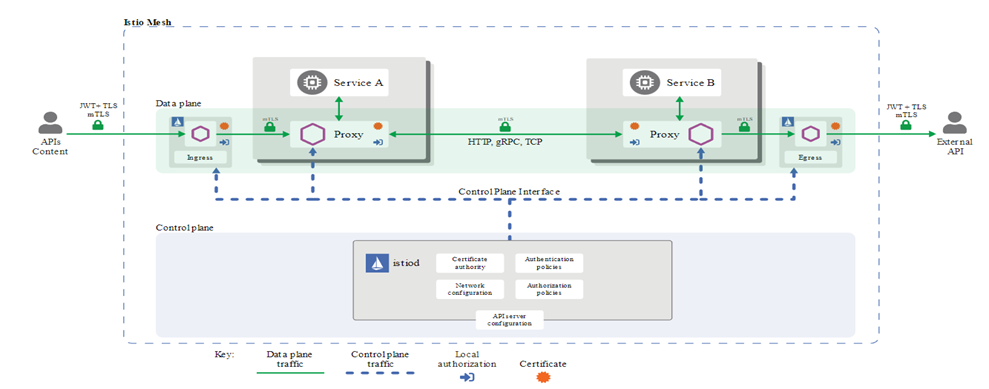}
    \caption{Istio architecture with various components and security controls. \textit{Source}: \cite{31Istio_Security}}
    \label{fig:6}
\end{figure}

Istio is an open-source service mesh tool that provides service-to-service communication for distributed applications in micro-service architecture \cite{30Istio_ServiceMesh}. Istio has two components, as shown in Figs. \ref{fig:5} and \ref{fig:6}:

\begin{itemize}
    \item Control Plane
    \item Data Plane
\end{itemize}

The data plane facilitates communication between the distributed services. Istio uses an Envoy proxy for each service deployed in the Kubernetes cluster or virtual machine. The control plane includes the management components to configure the routing, security policies, and other elements to support dynamic change as the environment requires \cite{30Istio_ServiceMesh}.

Istio service mesh provides authentication using certificate-based identities through the Istio agent running alongside the Envoy proxy. In addition, Istio delivers the capability to implement mTLS (Mutual Transport Layer Security). When the mTLS STRICT authentication mode is enforced, the Envoy proxy validates the certificates and establishes a secure channel for communicating the traffic.

Finally, the Istio authorization policy components enforce traffic authorization using the Istio agent. The network configuration component provides routing and other session control policies to manage the traffic. These policies can be applied at the resource, namespace, or global (service mesh) level  \cite{31Istio_Security}.

\section{Architecture Benefits}

The proposed architecture includes the benefits of the zero trust principles and micro-segmentation. The design can incorporate enhanced security tools and controls tailored to meet specific business and technical requirements. It emphasizes scalability and elasticity, providing a modular framework that facilitates the integration of new security tools and managing both cloud and on-premises workloads. Additionally, this design streamlines operational support. Each component can be scaled horizontally, such as by adding new cloud service providers or virtual networks, and vertically, such as by increasing the number of Kubernetes clusters in response to new data types or evolving business needs. 

\begin{itemize}
    \item Segmentation at multiple layers to provide granular controls
    \item Cluster-level segregation based on data types
    \item Layer 7 authentication and authorization for micro-services
    \item Layer 3/4 to control access at network and port
    \item Encryption in transit enforcement
    \item Centralized ingress and egress connection using SDP
    \item Segregation at the cloud virtual network layer
    \item Application-level segregation at the namespace level
    \item Traffic controls (routing, load balancing)
    \item Multi-layer traffic control
    \item Zero trust network model
    \item Multiple enforcement points
\end{itemize}

\section{Implementation and Demonstration}

\begin{figure}
    \centering
    \includegraphics[width=1\linewidth]{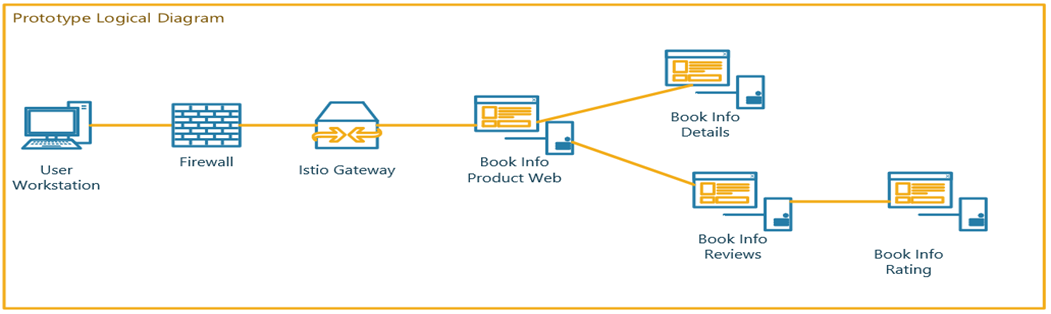}
    \caption{Prototype Logical Architecture Diagram}
    \label{fig:7}
\end{figure}

The proposed cloud micro-segmentation design is illustrated as the prototype implementation (Fig. \ref{fig:7}) to prove the various micro-segmentation design concepts highlighted in the previous section.

This prototype provides a feasibility analysis and demonstration of micro-segmentation security controls using open-source tools and technologies. The tools used in this prototype may be replaced with other open-source tools and technologies currently available or in the future for better performance, management, and operations as demanded by business and technical requirements.

In this prototype architecture, we have used the following tools and technologies:
\begin{itemize}
    \item  Virtual Servers using  Windows Hyper-VM
    \item Ubuntu 20.04.5 Operating System
    \item Kubernetes Cluster with control node and one worker node
    \item Calico as (CNI) container network interface for network security
    \item Istio for service mesh
    \item SonicWall TZ Series Next-Generation Firewall (NGFW) appliance
    \item User laptops for testing
    \item Networking components
\end{itemize}

The prototype deploys a Kubernetes-based micro-service architecture using standard foundational components such as virtual machines and operating systems. These can be customized or changed according to design requirements. To achieve zero trust security, the following layers of controls were deployed in the prototype design.

Perimeter Security:

\begin{itemize}
    \item Sonicwall Firewall Intrusion Prevention System
    \item Public Internet Inbound and Outbound access was limited using Sonicwall NGFW firewall.
    \item Sonicwall TZ Series Next-Generation Firewall (NGFW) appliance had SSL and IPsec VPN features, but they were not used for this prototype.
\end{itemize}

Network Security:
\begin{itemize}
    \item Kubernetes cluster in a private network to protect it from direct exposure to the Public Internet.
    \item Network ingress and egress traffic control using calico network and global network policies.
    \item mTLS (Mutual Transport Layer Security) for micro-services—Istio provides service-to-service communication over a secure channel using mTLS. Istio injects a proxy as a local sidecar service that allows traffic to intercept and initiate mTLS authentication and an encrypted traffic channel.
\end{itemize}

Application Security:

\begin{itemize}
    \item User Authentication: This prototype configures applications for testing for local user authentication.
    \item System authentication: Use of host-to-host mutual TLS (Transport Layer Security) for pod-level authentication
    \item Authorization: Istio provides authorization policies at service mesh, namespace, and workload levels. For this prototype, we deployed namespace-level authorizations.
\end{itemize}

Microservices Security:

\begin{itemize}
    \item Use of Kubernetes namespace to segregate the applications.
    \item Istio virtual service
    \item Istio Ingress and Egress Gateway
\end{itemize}

The following section outlines the deployment and testing of micro-segmentation implementation using Istio service mesh and mutual TLS implementation in the prototype environment.

Demonstration 1: Micro-segmentation using Istio service mesh.
This demonstration is focused on testing the Istio service mesh level micro-segmentation in the Kubernetes namespace within a cluster. By default, the Kubernetes cluster has no restrictions on controlling traffic. The Istio service mesh was deployed with the authorization policy to handle the connections between namespaces to enable micro-segmentation within the Kubernetes cluster.  The Kubernetes namespace contains two HTTP pods. These pods were enabled with Istio services mesh, and an authorization policy was configured to deny all traffic within the namespace to change the default allowed behavior.

Step 1: Create two pods in a Kubernetes namespace.

\begin{lstlisting}[style=bashstyle]
$ kubectl create ns httpsns-withistio
  kubectl label ns httpsns-withistio istio-injection=enabled
namespace/httpsns-withistio created
namespace/httpsns-withistio labeled
$
$ kubectl apply -f samples/httpbin/httpbin.yaml -n httpsns-withistio
serviceaccount/httpbin created
service/httpbin created
deployment.apps/httpbin created
\end{lstlisting}

\begin{lstlisting}[style=bashstyle]
$ kubectl apply -f samples/httpbin/httpbin.yaml -n httpsns-withistio
serviceaccount/httpbin created
service/httpbin created
deployment.apps/httpbin created
$
$ kubectl apply -f samples/httpbin/httpbin.yaml -n httpsns-withistio
  kubectl apply -f samples/sleep/sleep.yaml -n httpsns-withistio
serviceaccount/httpbin unchanged
service/httpbin unchanged
deployment.apps/httpbin unchanged
serviceaccount/sleep created
service/sleep created
deployment.apps/sleep created
\end{lstlisting}

Step 2: As no authorization policy is enabled, the connection between httpbin and sleep pod is allowed.

\begin{lstlisting}[style=bashstyle]
$ kubectl exec -it -n httpsns-withistio $SLEEP_POD -c sleep -- sh
$ 
$ curl -sI "http://httpbin:8000/ip" | grep HTTP
HTTP/1.1 200 OK
\end{lstlisting}

Step 3: An Istio authorization policy was configured and applied to restrict traffic with the namespace.

\begin{lstlisting}[style=bashstyle]
$ cat /tmp/istio-authpol.yml
apiVersion: security.istio.io/v1beta1
kind: AuthorizationPolicy
metadata:
  name: allow-nothing
  namespace: istio-system
spec:
  {}
\end{lstlisting}

\begin{lstlisting}[style=bashstyle]
$ kubectl apply -f /tmp/istio-authpol.yml
authorizationpolicy.security.istio.io/allow-nothing created
$
$ kubectl get authorizationpolicy -n istio-system
NAME            AGE
allow-nothing   29s
\end{lstlisting}

\begin{lstlisting}[style=bashstyle]
$ kubectl delete -n httpsns-withistio po/$HTTP_POD
pod "httpbin-ff5c9f7c-glkvw" deleted
$
$ kubectl get po -n httpsns-withistio
NAME                   READY   STATUS    RESTARTS   AGE
sleep-bc9998558-bmhrh   2/2     Running   0          9m6s
httpbin-ff5c9f7c-ctbgd  2/2     Running   0          19s
$
$ export HTTP_POD=httpbin-ff5c9f7c-ctbgd
$
$ kubectl exec -it -n httpsns-withistio $SLEEP_POD -c sleep -- sh
$ 
$ curl -sI "http://httpbin:8000/ip" | grep HTTP
HTTP/1.1 403 Forbidden
\end{lstlisting}

This demonstrates connection authentication using mTLS between pods in a namespace. The Istio service mesh makes it transparent, removes the additional load of encrypting the connection, and provides application authentication in the microservice architecture.

Demonstration 2: Mutual TLS (mTLS) deployment
When Kubernetes pods are enabled with Istio service mesh, envoy proxy is automatically deployed with pods, intercepts the connection, and applies configured policies.  By default, Istio allows mTLS authentication in a permissive way. This way, the default configuration will enable workloads to be moved gradually to a more secure environment as developers prepare the workload to run under a secured environment without breaking applications. 

Three sample HTTP applications and namespaces were created for this demonstration: 1. Foo, 2. Bar, and 3. Legacy. Foo and Bar were configured with mTLS permissive (default configuration) and strict mode, respectively, and Legacy was not configured to use mTLS. The below screenshots show the connection behavior without and with mTLS enabled in Istio. This demonstration will use a sample HTTP application to test the connection and access between three namespaces.

Communication pattern without Istio mTLS:

\begin{lstlisting}[style=bashstyle]
$ kubectl get peerauthentication --all-namespaces
No resources found
\end{lstlisting}

The first HTTP application namespace is ‘Foo’. It is deployed with an envoy proxy and is part of the Istio service mesh.

\begin{lstlisting}[style=bashstyle]
$ kubectl create ns foo
namespace/foo created
$
$ kubectl apply -f <(istioctl kube-inject -f samples/httpbin/httpbin.yaml) -n foo
serviceaccount/httpbin created
service/httpbin created
deployment.apps/httpbin created
$
$ kubectl apply -f <(istioctl kube-inject -f samples/sleep/sleep.yaml) -n foo
serviceaccount/sleep created
service/sleep created
deployment.apps/sleep created
\end{lstlisting}

\begin{lstlisting}[style=bashstyle]
$ kg all -n foo
NAME                            READY   STATUS    RESTARTS   AGE
pod/httpbin-655c7d749b-cs5kk    2/2     Running   0          2m30s
pod/sleep-7d8945695d-vp47n      2/2     Running   0          2m24s

NAME                 TYPE        CLUSTER-IP       EXTERNAL-IP   PORT(S)     AGE
service/httpbin      ClusterIP   10.43.46.198     <none>        8000/TCP    2m30s
service/sleep        ClusterIP   10.43.57.9       <none>        80/TCP      2m24s

NAME                         READY   UP-TO-DATE   AVAILABLE   AGE
deployment.apps/httpbin      1/1     1            1           2m30s
deployment.apps/sleep        1/1     1            1           2m24s

NAME                                   DESIRED   CURRENT   READY   AGE
replicaset.apps/httpbin-655c7d749b     1         1         1       2m30s
replicaset.apps/sleep-7d8945695d       1         1         1       2m24s
\end{lstlisting}

The second HTTP application namespace is named ‘Bar’ and is deployed with an envoy proxy as part of the Istio service mesh.

\begin{lstlisting}[style=bashstyle]
$ kubectl create ns bar
namespace/bar created
$
$ kubectl apply -f <(istioctl kube-inject -f samples/httpbin/httpbin.yaml) -n bar
serviceaccount/httpbin created
service/httpbin created
deployment.apps/httpbin created
$
$ kubectl apply -f <(istioctl kube-inject -f samples/sleep/sleep.yaml) -n bar
serviceaccount/sleep created
service/sleep created
deployment.apps/sleep created
\end{lstlisting}

Testing shows that access to the HTTP application from namespace Foo to Bar is successful.

 \begin{lstlisting}[style=bashstyle]
$ kubectl create ns bar
namespace/bar created
$
$ kubectl apply -f <(istioctl kube-inject -f samples/httpbin/httpbin.yaml) -n bar
serviceaccount/httpbin created
service/httpbin created
deployment.apps/httpbin created
$
$ kubectl apply -f <(istioctl kube-inject -f samples/sleep/sleep.yaml) -n bar
serviceaccount/sleep created
service/sleep created
deployment.apps/sleep created
\end{lstlisting}

The third HTTP application namespace is named ‘Legacy’ and deployed without an envoy proxy and is not part of the Istio service mesh. Since mTLS is not enforced and mandated by default in the Istio service mesh, access from Legacy to the Bar is allowed.

\begin{lstlisting}[style=bashstyle]
$ k exec -it -n legacy pod/sleep-bc9998558-kshxk -c sleep -- curl http://httpbin.bar:8000/ip -s -o /dev/null -w "sleep.legacy to httpbin.bar: %{http_code}\n"
sleep.legacy to httpbin.bar: 200
\end{lstlisting}
 
Communication pattern with Istio mTLS:
To test the connectivity with mTLS enforced within the service mesh, the Bar namespace is configured with mTLS STRICT mode. This mode only allows encrypted communication between pods and rejects connections that do not support mTLS.

\begin{lstlisting}[style=bashstyle]
$ kubectl apply -n bar -f - <<EOF
apiVersion: security.istio.io/v1beta1
kind: PeerAuthentication
metadata:
  name: default
spec:
  mtls:
    mode: STRICT
EOF
peerauthentication.security.istio.io/default created
\end{lstlisting}

Testing from namespace Foo to Bar is successful, as both namespaces are part of the Istio service mesh. The Bar namespace is configured strictly to use mTLS, whereas the Foo name is in the permissive namespace by default. mTLS Permissive mode allows connections with and without permissive mode.

\begin{lstlisting}[style=bashstyle]
$ k exec -it -n foo pod/sleep-7d8945695d-vp47n -c sleep -- curl http://httpbin.bar:8000/ip -s -o /dev/null -w "sleep.foo to httpbin.bar: %{http_code}\n"
sleep.foo to httpbin.bar: 200
\end{lstlisting}

Testing from namespace Legacy to Bar is unsuccessful as the Legacy namespace is not configured with the envoy proxy and does not support mTLS connections.

\begin{lstlisting}[style=bashstyle]
$ k exec -it -n legacy pod/sleep-bc9998558-kshxk -c sleep -- curl http://httpbin.bar:8000/ip -s -o /dev/null -w "sleep.legacy to httpbin.bar: %{http_code}\n"
sleep.legacy to httpbin.bar: 000
command terminated with exit code 56
\end{lstlisting}

This demonstration proves that once the authentication policy is enabled within the Istio service mesh, connections with mTLS are allowed, and access is denied if mTLS is not enabled. Istio authorization policy controls the traffic based on the policy configured using envoy proxy pods deployed with application service pods. 

\section{Conclusion and Future Work}

This paper presents a cloud-agnostic micro-segmented network architecture based on zero trust principles. The proposed architecture addresses several of the traditional network architecture issues. First, it offers an approach to implement next-generation network architecture using the cloud-native network, open-source tools, and support micro-service architecture. Second, it illustrates the various levels of segregation at network, resources, and application levels. Then, the architecture uses open-source network security and service mesh tools, such as Calico and Istio, to implement the authentication, authorization, and encryption using mutual TLS required to build a zero trust foundation. Third, it provides a consistent security posture to protect from internal, external, and lateral movement threats. Finally, the proposed architecture supports a multi-cloud strategy to build a secure network design based on the zero trust foundational building blocks to host modern applications architecture.
There are opportunities for future work in the operational aspects of the proposed design. These specific operational aspects may influence the change in the current approach for scalability and integration:
\begin{itemize}
\item Identity Governance, as every micro-service in service mesh, will use identity for authentication and mTLS.
\item Configuration monitoring and drift Management as operational authentication and authorization policies may be managed by DevOps staff. Hence, it will require operational controls, security guardrails, and monitoring to ensure security policies comply with the required standards and best practices.
\item Certification Management and Automation – Certificates are used for application authentication and encryption within service mesh, so issuance, revocation, and other aspects of the certificate lifecycle require adequate management controls.  
\item Processes and procedural complexities – Authentication and authorization policies are applied at multiple levels within the proposed design (resource, namespace, and global) and may be managed by DevOps staff. In a fast-paced environment, following the change management processes for every policy change may be impossible without impacting the velocity and agility to achieve the desired outcome.
\end{itemize}

\section*{Acknowledgment}

Grammarly was utilized in the writing process to assist with editing, spell-checking, and grammar enhancement. All content and ideas presented in this paper are our own.

\printbibliography

\end{document}